\documentclass[conference]{IEEEtran}

\usepackage{graphicx}
\usepackage{amsmath,amssymb,amsfonts}
\usepackage{multirow}
\usepackage{psfrag}
\usepackage{subfigure}
\usepackage{color}
\usepackage{soul}
\usepackage{xcolor}
\usepackage[hyphens]{url}
\usepackage[bookmarks=false]{hyperref}
\usepackage[hyphenbreaks]{breakurl}
\usepackage{lipsum}
\usepackage[ruled,vlined,linesnumbered]{algorithm2e}
\usepackage{mathrsfs}
\usepackage{pbox}
\usepackage{wrapfig}
\usepackage{flushend}

\DontPrintSemicolon

\makeatletter
\def\hlinewd#1{%
  \noalign{\ifnum0=`}\fi\hrule \@height #1 \futurelet
   \reserved@a\@xhline}
\makeatother

\graphicspath{{Figures/}}
\ifCLASSINFOpdf
\else
\fi

\begin{document}

\title{An Experimental Study on Direction Finding of Bluetooth 5.1: Indoor vs Outdoor}

\author{Pradeep~Sambu and Myounggyu~Won\\
Department of Computer Science\\
Connected Smart Sensor Systems (CS$^3$) Lab\\
University of Memphis, TN, United States\\
\{psambu, mwon\}@memphis.edu
}
%\author{Myounggyu~Won$^{1}$, Sayan~Sahu$^{1}$, Yunfan~Zhang$^{1}$, and Yongsoon Eun$^2$\\
%$^1$WENS Lab, South Dakota State University, Brookings, SD, United States\\
%$^2$CPS Global Center, Daegu Gyeongbuk Institute of Science and Technology, Daegu, South Korea\\
%\{myounggyu.won,sayan.sahu\}@sdstate.edu, yeun@dgist.ac.kr}%
%\thanks{Authors$^{1\dagger}$ equally contributed to this work.}%
%\thanks{Authors$^{1\dagger}$ equally contributed to this work.}

\markboth{Journal of \LaTeX\ Class Files,~Vol.~13, No.~9, September~2014}%
{Shell \MakeLowercase{\textit{et al.}}: Bare Demo of IEEEtran.cls for Journals}

\maketitle

\begin{abstract}
The Bluetooth Special Interest Group (Bluetooth SIG) introduced a new feature for highly accurate localization called the Direction Finding in the Bluetooth Core Specification 5.1. Since this new localization feature is relatively new, despite the significant interest of industry and academia in the accurate positioning of Bluetooth devices/tags, there are only a handful of experimental studies conducted to evaluate the performance of the new technology. Furthermore, these experimental works are constrained to only indoor environments or performed with hardware emulation of Bluetooth 5.1 via Universal Software Radio Peripherals (USRPs). In this paper, we perform an experimental study on the positioning accuracy of the direction finding using COTS Bluetooth 5.1 devices in booth indoor and outdoor environments to provide insights on the performance gap under these different experimental settings. Our results demonstrate that the average angular error in an outdoor environment is 0.28$^{\circ}$, significantly improving the angular error measured in an indoor environment by 73\%. It is also demonstrated that the average positioning accuracy measured in an outdoor environment is 22cm which is 39.7\% smaller than that measured in an indoor environment.
\end{abstract}


\IEEEpeerreviewmaketitle

\section{Introduction}
\label{sec:introduction}

The proliferation of localization services for various Internet of Things (IoT) applications such as smart health~\cite{baig2013smart}, disaster management~\cite{han2017disaster}, smart building~\cite{turgut2016indoor}, and surveillance~\cite{quigley2005target} necessitates highly accurate cm-level positioning systems~\cite{zafari2019survey}. Recently, the Bluetooth Special Interest Group (Bluetooth SIG) has launched a new feature on the highly accurate localization of Bluetooth devices called the Direction Finding in their Bluetooth Core Specification 5.1~\cite{bluetoothspec}. This new localization feature of Bluetooth 5.1 provides two different options for positioning a Bluetooth device namely Angle of Arrival
(AoA) and Angle of Departure (AoD). AoA allows a receiver with a multi-antenna array to identify the angular position of a transmitter based on the phase delay of the signal transmitted by the transmitter~\cite{bluetoothspec}. On the other hand, AoD allows the transmitting device with multiple antennas to transmit a radio signal that permits the receiver to determine the directional angle to the transmitter~\cite{bluetoothspec}.

The Direction Finding of Bluetooth 5.1 is a relatively new technology. As such, the industry has just started producing devices supporting this feature. As a result, there are only a handful of experimental studies conducted to evaluate the positioning accuracy of Direction Finding~\cite{cominelli2019dead}\cite{andersson2020evaluation} despite the significant prospects of the technology. One of the notable experimental study is performed to evaluate the performance of Direction Finding based on AoA~\cite{cominelli2019dead}. The authors demonstrate that even with a very simple hardware configuration for a receiver equipped with two antennas, sub-meter localization accuracy can be achieved. However, it is also noted in their paper that achieving a cm-level localization using Direction Finding is still challenging. We are aware of another latest experimental study performed at KTH Royal Institute of Technology that demonstrates the highly accurate positioning accuracy of Direction Finding~\cite{andersson2020evaluation}.

Although these state-of-the-art experimental studies present interesting and valuable results, we notice that two significant elements are missing. Firstly, the experiment is performed via hardware emulation based on USRPs (Ettus B 210 and Ettus N200~\cite{usrp}), possibly not representing the performance of Direction Finding accurately compared to conducting experiments with commercial off-the-shelf (COTS) Bluetooth devices equipped with an actual Bluetooth 5.1 chipset~\cite{cominelli2019dead}. Secondly, the experiment is performed focusing explicitly on an indoor environment~\cite{cominelli2019dead}\cite{andersson2020evaluation}. However, considering numerous outdoor Bluetooth applications~\cite{seymer2019secure}\cite{mackey2020smart}\cite{chien2020low}, it is worth to investigate the performance of Direction Finding in an outdoor environment and provide insights on the performance difference between these two heterogeneous environments.

To address these issues, in this paper, we present an experimental study on the angular and positioning accuracy of Direction Finding using COTS Bluetooth 5.1 devices. In particular, experiments are performed in both indoor and outdoor environments to provide novel insights on the performance of Direction Finding under different environments. Our results demonstrate that the average angular accuracy (0.28$^{\circ}$) in an outdoor environment is significantly improved by 73\% compared with that measured in an indoor environment. We also demonstrate that the average positioning error (22cm) in an outdoor environment is improved by 39.7\% compared with that measured in an indoor setting.

Overall, our contributions are summarized as follows.

\begin{itemize}
  \item We perform an experimental study to evaluate the angular and positioning accuracy of Direction Finding for Bluetooth 5.1 in both indoor and outdoor environments.
  \item Our experimental study is conducted using COTS Bluetooth 5.1 devices.
  \item We demonstrate the significant performance difference between indoor and outdoor settings in terms of the angular and positioning errors for Direction Finding.
\end{itemize}

This paper is organized as follows. A review of the Direction Finding technology is presented in Section~\ref{sec:background} to help readers better understand the results of this experimental study. We then describe the details of our experimental settings in Section~\ref{sec:implementation}. And then, the results and analysis are presented in Section~\ref{sec:results}. Finally, we conclude in Section~\ref{sec:conclusion}.

\section{Direction Finding of Bluetooth 5.1}
\label{sec:background}

The Bluetooth technology has been utilized for providing numerous localization applications for transportation~\cite{sawant2004using}, healthcare~\cite{laine2014mobile}, surveillance~\cite{kuxdorf2020passive}, and disaster management~\cite{yang2017disaster}. Especially, the Bluetooth Low Energy (BLE) beacon has been widely adopted and very successful~\cite{chawathe2008beacon}. Currently, there are over 130 million BLE beacons are being used all over the world. It is anticipated that more than hundreds of millions of them will be newly shipped every year within the next few years~\cite{bluetoothdirection}.

\subsection{Basic Concept}

To support such explosively increasing demands for the Bluetooth localization services, the Bluetooth Special Interest Group has introduced a new localization feature called Direction Finding in the Bluetooth Core specification 5.1~\cite{bluetoothdirection}. To achieve highly accurate positioning, Direction Finding improves the host controller interface (HCI) to support specialized hardware built with a multi-antenna array so that Bluetooth devices can determine the direction of the received radio signal. Additionally, it allows data acquired from the received signal to be made available to the higher layers of the software stack to facilitate calculation of the direction of the signal.

\begin{figure}[h]
\centering
\includegraphics[width=.95\columnwidth]{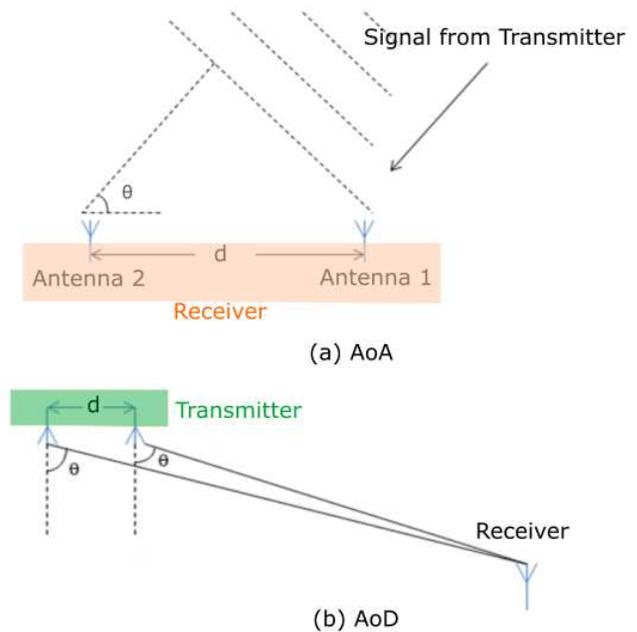}
\caption {(a) Angle of Arrival (AoA) and (b) Angle of Departure (AoD). Both methods are supported by Bluetooth 5.1 to allow a a device to determine the direction of the received signal.}
\label{fig:aoaaod}
\end{figure}

The new feature of Bluetooth 5.1 provides two options: Angle of Arrival (AoA) and Angle of Departure (AoD). In AoA, the receiver is equipped with an antenna array and performs computation of the direction of the received signal as illustrated in Fig.~\ref{fig:aoaaod}(a). On the other hand, in AoD, the transmitter has an antenna array, and the receiving device has a single antenna (Fig.~\ref{fig:aoaaod}(b)). The receiver calculates the direction of the signal based on the phase difference of received signals from the transmitter equipped with the antenna array. Specifically, given the distance between two antennas denoted by $d$, the wave length of the signal ($\lambda$), and the phase difference ($\psi$), a simple trigonometry yields the angle of the received signal, \emph{i.e.,} $\theta = arccos(\frac{\psi \lambda}{2\pi d})$ for AoA, and $\theta = arcsin(\frac{\psi \lambda}{2\pi d})$ for AoD.

\subsection{Direction Finding Signal}

\begin{figure}[h]
\centering
\includegraphics[width=.98\columnwidth]{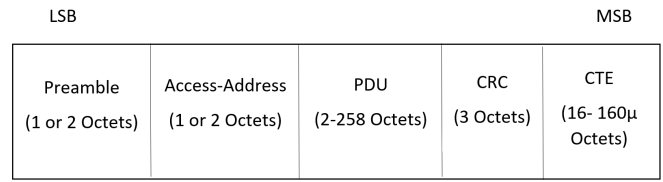}
\caption {The signal structure for Direction Finding. This specially designed signal is the essential component of Direction Finding.}
\label{fig:packet_struct}
\end{figure}

To enable accurate localization, Direction Finding exploits specially designed direction finding signals~\cite{bluetoothdirection}. Fig.~\ref{fig:packet_struct} displays the structure of the direction finding signal. A novel link layer Protocol Data Unit (PDU) is provided to support the localization service between two connected devices. Direction Finding also allows to use an existing PDU to support localization service for connectionless devices. The Constant Tone Extension (CTE) field at the end of the signal is a series of symbols representing binary 1 that is used for IQ sampling. Direction Finding allows the higher layer to configure the number of symbols in the CTE field corresponding to the hardware of the Bluetooth device that may require different data and time for IQ sampling. The Cyclic Redundancy Check (CRC)~\cite{tsimbalo2016crc} field of the signal is used for error correction. Additionally, Direction Finding allows the PDU field to contain a Message Integrity (MIC) field if the connection between devices is encrypted and authenticated.

\begin{figure}[h]
\centering
\includegraphics[width=0.99\columnwidth]{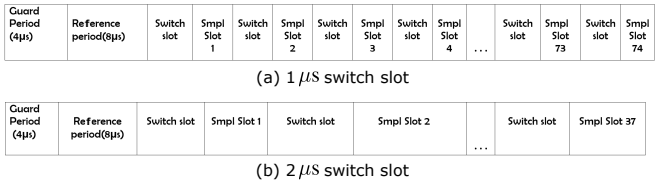}
\caption {The CTE field of a direction finding signal with (a) 1$\mu$s switch slot and (b) 2$\mu$s switch slot. The 1$\mu$s slot is provided as an option while the 2$\mu$s is mandatory.}
\label{fig:bluetooth_packet}
\end{figure}

The CTE field is divided into a 4$\mu$s guard period, an 8$\mu$s reference period, and a sequence of switch slots and sample slots~\cite{bluetoothdirection} (Fig.~\ref{fig:bluetooth_packet}). The guard period is used to ensure that two subsequent signals do not interfere with each other by allowing some time gap between them. Eight IQ samples are collected during the 8$\mu$s reference period at 1$\mu$s intervals. These IQ samples are used to estimate the information about the received signal, \emph{i.e.,} the wave length and frequency to calculate the angle more accurately. Note that during the reference period, antenna switching does not occur. A sequence of switch slots and time slots follow the reference period. During the sample slot, IQ sampling is performed, while in the switch slot, antenna switching is performed. The time period for the sampling slot as well as the switch slot can be 1$\mu$s or 2$\mu$s. The 1$\mu$s slot is optional while the 2$\mu$s slot is mandatory.

%\begin{wrapfigure}{r}{0.5\columnwidth}
%\vspace{-15pt}
%  \begin{center}
%    \includegraphics[width=\linewidth]{triangulation}
%    \vspace{-15pt}
%    \caption{title. and text.  \label{fig:triangulation}}
%  \end{center}
%  \vspace{-10pt}
%\end{wrapfigure}

\section{Experimental Setup}
\label{sec:implementation}

\begin{figure}[h]
\centering
\includegraphics[width=.97\columnwidth]{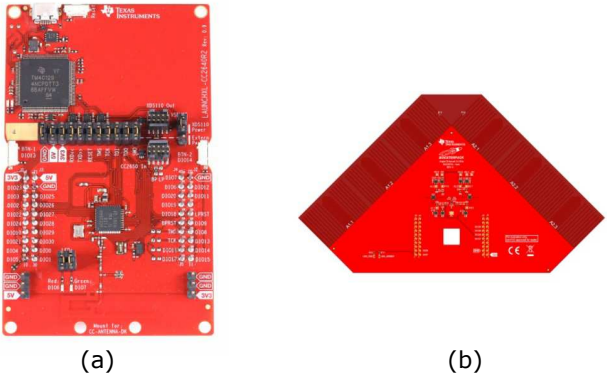}
\caption {(a) TI Bluetooth transceiver (CC26X2R LaunchPad) and (b) antenna array (BOOSTXL-AOA). The receiver for AoA is created by integrating the antenna array with the LaunchPad.}
\label{fig:devices}
\end{figure}

There are several companies that offer a development kit for implementing the Direction Finding of Bluetooth 5.1. Some of those companies are Silicon Labs~\cite{silabs}, Laird Connectivity~\cite{laird} and Texas Instruments (TI)~\cite{titool}. We select TI's product for conducting this experimental study as they provide not only a Bluetooth 5.1 transceiver but also the antenna array package as shown in Fig.~\ref{fig:devices}. Furthermore, the source code of the Bluetooth 5.1 protocol stack is fully available as well as a well-established integrated development environment (IDE) tools with rich libraries to facilitate the application development process~\cite{titool}.

We use AoA in this experimental study. As such, a single transmitter and two receivers integrated with a multi-antenna array are required to localize the position of the transmitter. The antenna array (Fig.~\ref{fig:devices}(b)) consists of 6 antennas with 3 antenna switches~\cite{titool}. The Bluetooth transceiver as shown in Fig.~\ref{fig:devices}(a) is used as the receiver for AoA by integrating with the antenna array module. Specifically, the hardware configuration of the transceiver is modified by relocating a micro capacitor on the transceiver board to connect with the antenna array via JSC cable.  We use a hot air de-soldering rework station~\cite{hotair} to carefully reposition the micro capacitor (due to the extremely small size) to integrate the antenna array with the transceiver. In contrast to the receiver, the transmitter does not require any hardware reconfiguration, \emph{i.e.,} the CC26X2R LaunchPad is simply used as the transmitter. The firmware of the transmitter is designed to allow the transmitter to transmit a Direction Finding signal periodically. The receiver is programmed to monitor the channel to receive Direction Finding signals. Each Bluetooth device (\emph{i.e.,} the two receivers and a transmitter) is connected to a laptop equipped with a Intel 2.3GHz i7 CPU and 16GB of RAM running on 64-bit Windows 10 Operating System to receive power via USB, compute the angle of arrival, and log the calculated angle.

\begin{figure}[h]
\centering
\includegraphics[width=.97\columnwidth]{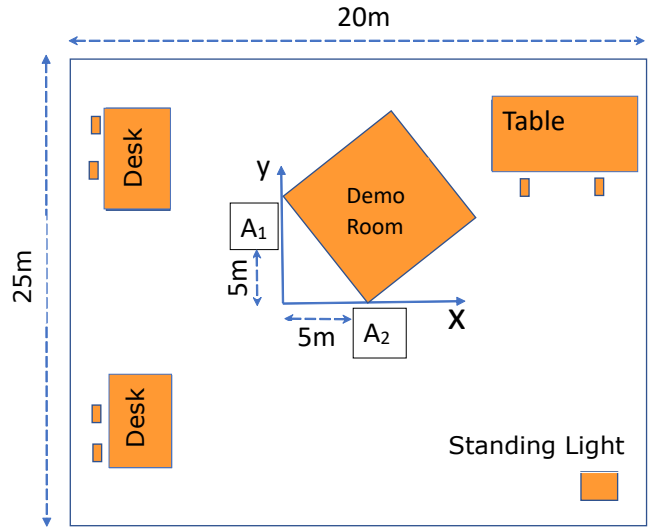}
\caption {A diagram of the indoor environment. The obstructions (\emph{i.e.,} desks, a table, and a standing light) could possibly create the multi-path effect, degrading the positioning performance.}
\label{fig:indoor_setup}
\end{figure}

\begin{wrapfigure}{r}{0.5\columnwidth}
\vspace{0pt}
  \begin{center}
    \includegraphics[width=\linewidth]{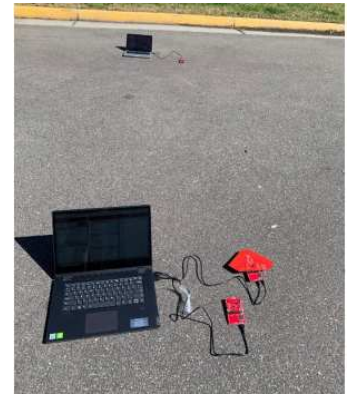}
    \vspace{-15pt}
    \caption{A screenshot of the outdoor experimental environment. \label{fig:indoor_outdoor_photo}}
  %\caption{A gull}
  \end{center}
  \vspace{-10pt}
\end{wrapfigure}

\begin{figure*}[t]
\centering
\includegraphics[width=.99\textwidth]{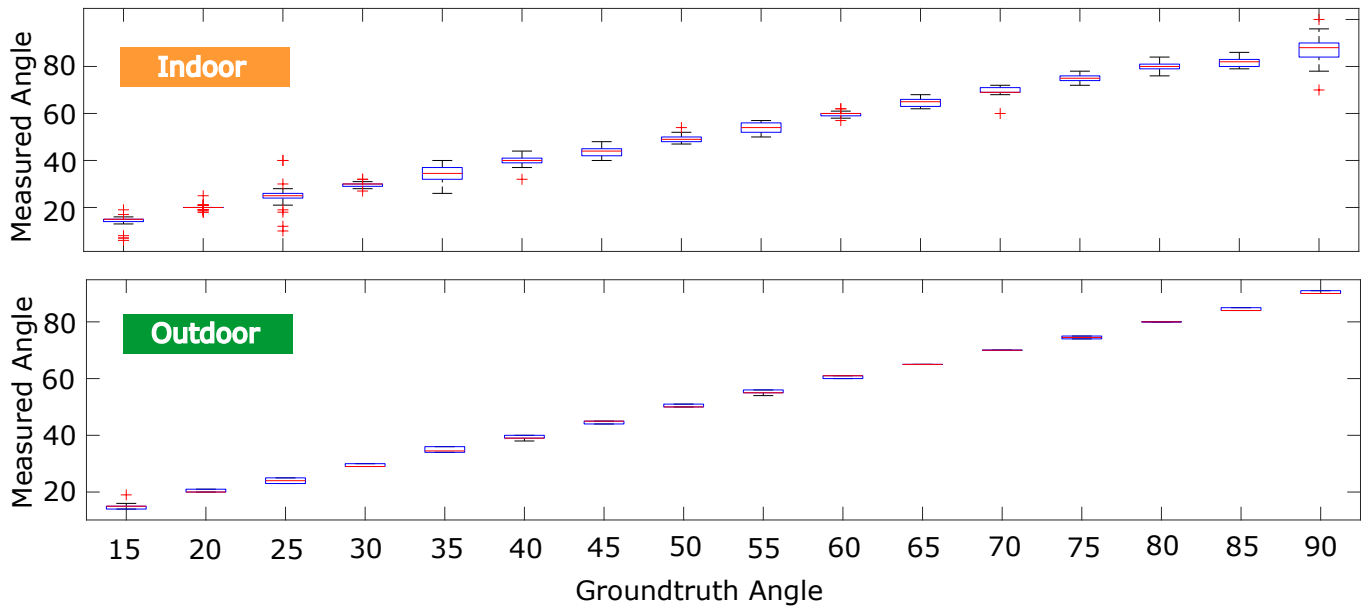}
\caption {The angular accuracy for indoor and outdoor environments. The average angular accuracy for the outdoor environment is 73\% smaller than that measured in the indoor environment.}
\label{fig:angle_results}
\end{figure*}

We perform experiments in both indoor and outdoor environments to evaluate the performance of Direction Finding. Fig.~\ref{fig:indoor_setup} exhibits a diagram of the indoor environment. Two receivers with antenna arrays denoted by $A_1$ and $A_2$ respectively are deployed in the middle of the room. The transmitter is placed at a predefined position within the demo area corresponding to a ground truth angle (for evaluating the angular accuracy) or a ground truth position (for evaluating the positioning accuracy) so that we can measure the  the angular and positioning error. As shown in Fig.~\ref{fig:indoor_setup}, there are some obstructions in the room (\emph{i.e.,} desks, table, and standing light) that could possibly create the multipath effect and degrade the positioning performance~\cite{sen2013avoiding}. The experiment is also performed in an outdoor environment. An open space with minimal obstruction is selected for the outdoor experiments (Fig.~\ref{fig:indoor_outdoor_photo}). The same experimental setting as that for the indoor environment is used for the outdoor environment, \emph{i.e.,} the same anchor locations and predefined positions for the transmitter corresponding to the ground truth angles and positions, to compare the performance of Direction Finding in two different environments.

\section{Results}
\label{sec:results}

The experimental results are presented in two metrics: the angular error and 2D positioning error. The angular error is represented as the difference between the measured angle and the ground truth angle. Similarly, the 2D positioning error is defined as the Euclidean distance between the measured position and the ground truth position. In measuring the positioning error, two anchors (\emph{i.e.,} receivers) are deployed to obtain an angle from each anchor which is used to perform triangulation to get the position of the transmitter.

\subsection{Angular Accuracy}
\label{subsec:indoor}

The angle $\theta$ between the direction of the signal propagation and the axis of the antenna array is measured in both indoor and outdoor environments. The transmitter is placed at a predefined position corresponding to the ground truth angle. The ground truth angles between $15^{\circ}$ and $90^{\circ}$ are considered with an increment of $5^{\circ}$. For each ground truth angle, we collect 30 angle measurements. The reason why the angles smaller than $15^{\circ}$ are not considered in our experiment is due to the observation noted in a state-of-the-art experimental study that when the direction of signal propagation is very close to the axis of the antenna array, the phase-delay can be random~\cite{cominelli2019dead}.

Fig.~\ref{fig:angle_results} depicts the blox plot of the angular accuracy for both indoor and outdoor environments. We observe that overall Direction Finding provides very accurate angular measurements in both indoor and outdoor environments. An interesting observation is that the standard deviation for smaller angles close to 0$^{\circ}$ is higher than larger angles $> 35^{\circ}$. Such a phenomenon is more evident in the results obtained from the indoor environment than the outdoor environment. A possible reason according to the observation made by Cominelli \emph{et al.}~\cite{cominelli2019dead} is because of the higher chance of getting phase-delay to wrap around when the angle is close to 0~\cite{cominelli2019dead} even though we excluded smaller angles smaller than $15^{\circ}$.

Now we compare the angular error of indoor environment with that of the outdoor environment. Overall, The average angular error of all angles for the indoor environment is 1.83$^{\circ}$, and the standard deviation is 1.1$^{\circ}$. The average angular error for the outdoor environment is 0.48$^{\circ}$ with the standard deviation of 0.28$^{\circ}$, which is 73\% smaller than that measured in the indoor environment. The results demonstrate that the Direction Finding of Bluetooth 5.1 performs significantly better in an outdoor environment. The results also indicate that the Direction Finding of Bluetooth 5.1 is affected by the multi-path effect significantly, which is one of the major sources of errors for indoor localization~\cite{sen2013avoiding}.

\subsection{Positioning Accuracy}

\begin{wrapfigure}{r}{0.5\columnwidth}
\vspace{-15pt}
  \begin{center}
    \includegraphics[width=\linewidth]{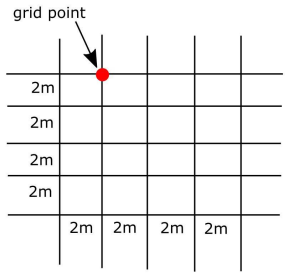}
    \vspace{-15pt}
    \caption{The 5 by 5 grid used to define the ground truth positions to measure the positioning accuracy.  \label{fig:grid}}
  %\caption{A gull}
  \end{center}
  \vspace{-10pt}
\end{wrapfigure}

Based on the angles obtained from two anchors/receivers, we perform triangulation to obtain a 2D position of the transmitter. We create a 5 by 5 grid region (\emph{i.e.,} 25 grid points) with the distance between grid points being 2m. Fig.~\ref{fig:grid} illustrates the grid region. A 10m by 10m space is secured to create the grid region in an indoor environment (Fig.~\ref{fig:indoor_setup}). For each grid point considered as the ground truth position, we repeat measurements of the 2D location of the transmitter 30 times. The same grid region is created in the outdoor environment.

\begin{figure}[h]
\centering
\includegraphics[width=.97\columnwidth]{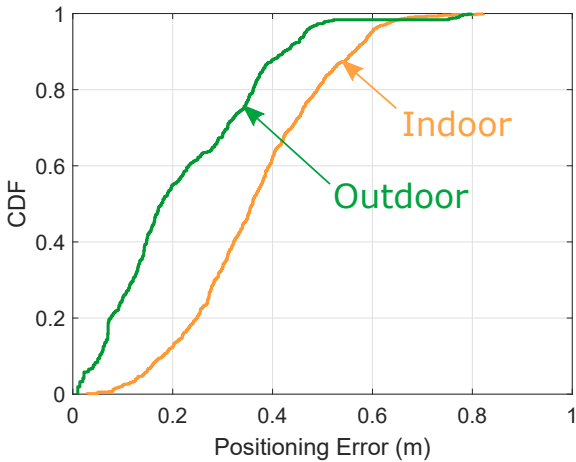}
\caption {The CDF graph of the positioning error for indoor and outdoor environments. The positioning error for the outdoor environment is 39.7\% smaller than that of the indoor environment.}
\label{fig:positioning_results}
\end{figure}

Fig.~\ref{fig:positioning_results} displays the cumulative distribution function (CDF) graph of the positioning error for both indoor and outdoor environments. The results demonstrate that overall the positioning accuracy for the outdoor environment is much higher than that for the indoor environment, which is somewhat expected as we already obtained better angular accuracy for the outdoor environment which is used to calculate the position. More precisely, the average positioning errors for the indoor and outdoor environments are 36.5cm and 22cm, respectively demonstrating that the positioning accuracy measured in the outdoor environment is smaller by 39.7\% compared with that for the indoor environment.

\section{Conclusion}
\label{sec:conclusion}

We have presented an experimental study on the angular and positioning accuracy for Direction Finding of Bluetooth 5.1. The experiment was conducted using COTS Bluetooth 5.1 devices in both indoor and outdoor environments, providing insights on the performance difference under these environmental settings for Direction Finding. Our results indicate that the Direction Finding performs significantly better in an outdoor environment by 73\% and 39.7\% in terms of the angular error and the positioning error, respectively. We expect that our experimental results will be useful information for academia and industry that plan to deploy Bluetooth localization applications using Direction Finding.

This research warrants a number of interesting future works worth to investigate. First, the impact of the distance between the receiver and transmitter can be studied as the signal strength (which generally weakens as the distance increases) may influence the positioning accuracy. Second, as there are many other COTS Bluetooth 5.1 devices currently available, it would be valuable to perform experiments with other Bluetooth devices and provide benchmark results. Third, the performance of Direction Finding in other environments can be studied such as an underwater environment. Finally, another interesting research problem is to design an effective localization algorithm specifically for the Direction Finding of Bluetooth 5.1. to improve the positioning accuracy in an indoor environment.

\bibliographystyle{IEEEtran}
\bibliography{mybibfile}

\end{document}